\documentstyle[12pt,aasms4]{article}

\def\be{\begin{equation}}
\def\ee{\end{equation}}

\def\eq#1{equation~(\ref{eq:#1})}

\def\today{\ifcase\month\or January\or February\or March\or April\or
May\or
June\or July\or August\or September\or October\or November\or
December\fi
  \space\number\day, \number\year}

\def\cl{\centerline}
\def\etal{{\it et al.\ }}

\def\eg{{e.g.}}

\def\ltsima{$\; \buildrel < \over \sim \;$}
\def\lsim{\lower.5ex\hbox{\ltsima}}
\def\gtsima{$\; \buildrel > \over \sim \;$}
\def\gsim{\lower.5ex\hbox{\gtsima}}
\def\ga{\mathrel{\hbox{\rlap{\hbox{\lower4pt\hbox{$\sim$}}}\hbox{$>$}}}}
\def\la{\mathrel{\hbox{\rlap{\hbox{\lower4pt\hbox{$\sim$}}}\hbox{$<$}}}}

\def\ifm#1{\relax\ifmmode#1\else$\mathsurround=0pt #1$\fi}
\def\kms{\,{\rm km\,s\ifm{^{-1}}}}

\def\la{\langle}

\def\pmb#1{\setbox0=\hbox{#1}%
 \kern-.025em\copy0\kern-\wd0
 \kern.05em\copy0\kern-\wd0
 \kern-.025em\raise.0433em\box0}

\def\vr{\pmb{$r$}}
\def\vv{\pmb{$v$}}

\def\vnabla{\pmb{$\nabla$}}
\def\div{\vnabla\!\cdot\!}

\def\divv{\div\vv}

\def \vr {{ \bf r}}
\def \br {{ \bf r}}
\def \bv {{ \bf v}}
\def \vv {{ \bf v}}

\def \bx {{ \bf x}}

\def \WF {^{\rm WF}}
\def \CR {^{\rm CR}}

\def\rd{{\rm d}}

\def \Omeg {\Omega_0}

\def \r0p{ r{_0^\prime}}
\def \prior{{\it prior}}
\def \posterior{{\it posterior}}
\def \m3{{\rm Mark III}}

\def\uo{u^o}

\begin{document}

\title{WIENER RECONSTRUCTION OF LARGE-SCALE STRUCTURE FROM PECULIAR
VELOCITIES}

\author{Saleem Zaroubi\altaffilmark{1,2}, 
        Yehuda Hoffman \altaffilmark{1}, 
        and Avishai Dekel \altaffilmark{1} }


\altaffiltext{1}{Racah Institute of Physics, The Hebrew University,
Jerusalem 91904, Israel}
\altaffiltext{2}{Max Planck-Institut f\"ur Astrophysik, 
                 Karl-Schwartzschild-Strasse 1,
                 65748 Garching, Germany}



\begin{abstract}

We present an alternative, Bayesian method for large-scale reconstruction
from observed peculiar velocity data. The method stresses a rigorous
treatment of the random errors and it allows extrapolation into poorly
sampled regions in real space or in k-space.
A likelihood analysis is used to determine the fluctuation power spectrum,
followed by a Wiener Filter (WF) analysis to obtain the minimum-variance
mean fields of velocity and mass density. Constrained Realizations (CR)
are then used to sample the statistical scatter about the WF mean field.
The method is tested using mock catalogs based on a simulation that
mimics the \m3\ data.  With low-resolution Gaussian smoothing of radius
$1200\kms$, the reconstruction is of high signal-to-noise
ratio ($S/N$) in a relatively large volume, with small
variance about the mean field.  The high resolution reconstruction,
of $500\kms$ smoothing, is of reasonable $S/N$ only in limited nearby
regions, where interesting new substructure is resolved.
The WF/CR method is applied as a demonstration to the Mark III data.
The main reconstructed structures are consistent with those extracted by
the POTENT method.  A comparison with the structures in the distribution
of IRAS 1.2Jy galaxies yields a general agreement.
The reconstructed velocity field is decomposed into its divergent
and tidal components relative to a 
cube of $\pm 8000 \kms$ centered on the Local Group.
The divergent component is very similar to the velocity field predicted
from the distribution of IRAS galaxies.
The tidal component is dominated by a bulk flow of $194\pm 32\kms$
towards the general direction of the Shapley concentration,
and it also indicates a significant quadrupole.

\end{abstract}


\vfill \eject

\section{INTRODUCTION}
\label{sec:intro}

The analysis of observed peculiar velocities plays a major role in current
research in cosmology and the formation of structure in the universe
(for reviews: Dekel 1994; Strauss \& Willick 1995; Willick 1998; Dekel 1998).
On scales large enough, where the
dynamics is dominated by gravity and the deviations from homogeneity
are small, the observed velocity field reflects the
dynamical evolution of structure and the total underlying mass distribution,
seen and unseen.  It can serve to measure cosmological parameters
such as the mean mass density, $\Omega$, and the comparison
with the distribution of luminous matter can shed light on the
``biasing" relation between galaxies and mass and thus on the process
of galaxy formation.  One way to achieve these goals is by {\it reconstruction}
of the full dynamical structure in the local cosmological neighborhood.
This paper presents a method for reconstruction
from the incomplete information provided by the observed radial velocities.
Complementary reconstruction can be done from whole-sky redshift surveys
(\eg, Strauss \etal 199x; Kolatt \etal 1996; 
Fisher \etal 1995; Bistolas \& and Hoffman 1998;
Narayanan \& Weinberg 1998).

This reconstruction from peculiar velocities faces several difficulties.
First, the distance measurements carry large random errors, including
intrinsic scatter in the distance indicator and measurement errors,
which grow in proportion to the distance from the observer and thus
become severe at large distances.
Further non-trivial errors are introduced by the nonuniform sampling
of the galaxies that serve as velocity tracers.  In particular, the Galactic
disk obscures an appreciable fraction of the sky, creating a significant
``Zone of Avoidance" (ZoA) of at least $40\%$ of the sky.
When translated to an underlying smoothed field, these errors give
rise to severe systematic biases as well
(\eg, Willick \etal\ 1995; 1996; 1997 for a detailed analysis of the errors).

The standard method of reconstruction from observed peculiar velocities
is the POTENT method (Bertschinger \& Dekel 1989; Dekel, Bertschinger
\& Faber 1990; recent versions described in Dekel 1994; Dekel 1998;
Dekel \etal 1998).
It has been applied, for example, to the Mark III data (Willick \etal\ 1995,
1996, 1997) with Gaussian
smoothing of $1200\kms$ within a volume of typical radius $\sim 6000\kms$.
In POTENT, the smoothed three-dimensional velocity field and the underlying
density field are recovered using the specific properties of mildly-nonlinear
gravitating flows, in particular their irrotationality.
The main effort in the POTENT method is directed towards minimizing the
biases in the smoothed velocity field. An attempt is made to weight each
data point properly by the inverse square of the random error, but this
weighting is eventually altered in a simultaneous attempt to also minimize
the systematic errors. The treatment of random errors in POTENT is thus
not optimal.

The present paper presents an alternative method of reconstruction,
which stresses the rigorous treatment of the random errors and allows
powerful extrapolation into poorly sampled regions.

The method presented here is Bayesian, where the \posterior\ probability
of a model is computed given the data and some prior assumptions about
the system.  In our case, the Bayesian \prior\ is formulated within the
standard model of cosmology which assumes that structure has emerged from
small-amplitude early fluctuations that constitute a Gaussian random field.
The statistical properties of a Gaussian field are completely determined
by the two-point correlation function or its Fourier conjugate, the power
spectrum.  The Bayesian formulation of the problem thus involves determining
the power spectrum and the smoothed density fluctuation field given the
data and the assumption of Gaussianity.
Zaroubi \etal (1995) have worked out such a method for the general
case of noisy and sparse observations of Gaussian fields. The solution found
consists of a maximum likelihood (hereafter MaxLike) estimation of the power
spectrum and subsequent Wiener filter (hereafter WF) reconstruction of
the underlying field for the MaxLike power spectrum.  Zaroubi \etal (1997)
have performed the first step by computing the MaxLike power spectrum
from the \m3\ data.
In the present paper, given this most likely power spectrum, we estimate
the underlying mean density and velocity fields via WF.

The Wiener Filter provides an optimal estimator of the underlying field in
the sense of a
minimum-variance solution given the data and an assumed \prior\ model (Wiener
1949; Press \etal\ 1992). The \prior\ defines the data auto-correlation
and the data-field cross-correlation matrices. In the case where the data is
drawn from a random Gaussian field, the WF estimator coincides with the
conditional mean field and with the most probable configuration given the
data. In the case of Gaussian fields where quadratic entropy can be assigned,
the WF also coincides with the maximum entropy solution (see Zaroubi
\etal 1995).

The common use of WF is for straightforward noise suppression, but
here it is generalized to achieve two further goals:
to reconstruct the density field from the observed radial velocities and to
interpolate or extrapolate the reconstruction to regions of poor sampling.
The latter can be done in real configuration space, to interpolate into
the ZoA or to extrapolate into large distances, or it can be used in
a reciprocal space, such as Fourier space or spherical-harmonic space,
to increase the resolution of the data.

The Wiener Filter multiplying the data to obtain the estimator is
schematically $P_k/(P_k+\sigma^2)$ (where $P_k$ is the power spectrum
and $\sigma$ is the error).  This means that the filter attenuates the
estimator to zero in regions where the noise dominates. The reconstructed
mean field is thus statistically inhomogeneous.
In order to recover statistical homogeneity we produce constrained
realizations (CR), in which random realizations of the residual from
the mean are generated such that they are statistically consistent both
with the data and the \prior\ model (Hoffman and Ribak 1991; see also
Bertschinger 1987).  In regions (in real or reciprocal space) dominated
by good quality data, the CRs are dominated by the data, while in the
limit of no data the realizations are practically unconstrained.

The WF/CR approach has been already applied to the IRAS two-dimensional galaxy
distribution (Lahav \etal\ 1994), the IRAS three-dimensional redshift
distortions (Fisher \etal\ 1995, Webster, Lahav and Fisher 1997),
the velocity potential out of POTENT (Ganon and Hoffman 1993),
the COBE/DMR cosmic microwave background mapping (Bunn \etal\ 1994)
and the COBE/DMR Jointly with the Tenerife data (Bunn, Hoffman and Silk 1996).
A nonlinear extension of the basic linear WF/CR has been applied to
the IRAS 1.2Jy catalog (Kolatt \etal\ 1996; Bistolas and Hoffman 1997).
The WF/CR is adjusted here to deal with peculiar velocities, and
is applied to the \m3\ catalog of peculiar velocities.

The Bayesian approach to velocity analysis was pioneered by Kaiser \& Stebbins
(1991). Their specific representation enabled only a limited dynamical
range in the reconstructed field; this was improved in Stebbins (1993)
by employing the more compact representation of Hoffman \& Ribak (1991).
The analysis presented here improves on these early works in three major ways:
(i) by using the much extended \m3\ data;
(ii) by using the MaxLike estimation of the \prior\ model;
and (iii) by employing CRs to sample the residual from the mean WF field.

The outline of this paper is as follows.
The method is presented in \S~\ref{sec:method}.
The data and the prior model are summarized in \S~\ref{sec:data}.
The method is tested using artificial data based on simulations in
 \S~\ref{sec:test}.
The results from the Mark III catalog are described in terms of
maps in \S~\ref{sec:results}.
Our results are discussed and the conclusions are summarized in
 \S~\ref{sec:conclusion}.

\section{WIENER FILTER AND CONSTRAINED REALIZATIONS}
\label{sec:method}

The general application of the WF/CR method to the reconstruction of
large-scale structure is described in Zaroubi \etal (1995),
where the theoretical foundation is discussed in relation with other
methods of estimation, such as Maximum Entropy.
Here we limit the description to the actual application of the method
to the case of radial velocity data.

The data for the WF/CR analysis are given as a set of observed radial
peculiar velocities $\uo_i$ sampled at
positions $\vr_i$ with estimated errors $\epsilon_i$ that are assumed to
be uncorrelated.  The peculiar velocities are assumed to be corrected for
systematic errors such as Malmquist bias.
The observed velocities are thus related to the true underlying velocity
field $\bv(\br)$, or its radial component $u_i$ at $\vr_i$, via
\be
\uo_i = \bv(\br_i) \cdot \hat \br_i  + \epsilon_i
\equiv u_i + \epsilon_i .
\label{eq:eps}
\ee

We assume that the peculiar velocity field $\vv(\vr)$
and the density fluctuation
field $\delta(\vr)$ are related via linear gravitational-instability theory,
$\delta = f(\Omega)^{-1} \divv$, where $f(\Omega)\approx\Omega^{0.6}$
and $\Omega$ is the mean universal density parameter.
Under the assumption of a specific theoretical prior for the power
spectrum $P(k)$ of the underlying density field,
we can write the WF minimum-variance estimator of the fields as
\be
\bv\WF(\br) = \Bigl < \bv(\br) \uo_i \Bigr >  \Bigl < \uo_i \uo_j \Bigr >
^{-1}        \uo_j
\label{eq:WFv}
\ee
and
\be
\delta\WF(\br) = \Bigl < \delta(\br) \uo_i \Bigr >  \Bigl < \uo_i \uo_j
\Bigr > ^{-1}    \uo_j  .
\label{eq:WFd}
\ee
In these equations $ \Bigl < ... \Bigr > $ denotes an ensemble average,
and we shall explain below how each term is calculated from the data
and the assumed power spectrum.
%
Note that in linear theory the WF reconstruction of the velocity and
density fields is equivalent to first reconstructing one of those
fields and then solving the Poisson equation for the other.

The formalism of constrained realizations allows us to create
typical realizations of the residual from the WF mean field
(Hoffman and Ribak 1991).  The method is based on creating random
realizations, $\tilde\delta(\br)$ and $\tilde\bv(\br)$,
of the underlying fields that obey the assumed power spectrum and
linear theory, and a proper set of random errors $\tilde\epsilon_i$.
The velocity random realization is then ``observed" like the actual data to
yield a mock velocity dataset $\tilde \uo_i$.  Constrained realizations
of the dynamical fields are then obtained by
\be
\bv \CR(\br) = \tilde \bv(\br) + \Bigl < \bv(\br) \uo_i \Bigr >
\Bigl < \uo_i \uo_j \Bigr > ^{-1} \bigl( \uo_j -\tilde \uo_j \bigr)
\label{eq:CRv}
\ee
and
\be
\delta \CR(\br) =
\tilde \delta(\br) +  \Bigl < \delta(\br) \uo_i \Bigr >  \Bigl < \uo_i \uo_j
\Bigr > ^{-1} \bigl( \uo_j -\tilde \uo_j \bigr) .
\label{eq:CRd}
\ee

The two types of covariance matrices in the above equations are computed
within the framework of linear theory as follows.
The covariance matrix of the data, to be inverted for the above
equations, can be written based on \eq{eps} as follows:
\be
R_{ij} \equiv
\Bigl < \uo_i \uo_j  \Bigr > = \Bigl < u_i u_j \Bigr > + \Bigl
                           < \epsilon_i \epsilon_j \Bigr >
   =  \hat \br_i \Bigl < \bv(\br_i) \bv(\br_j)  \Bigr > \hat \br_j
+ \sigma{^2_i}\delta_{ij} .
\label{eq:Rij}
\ee
The last term, involving the Dirac delta function $\delta_{ij}$,
is the diagonal matrix of error covariance.
The velocity covariance tensor that enters this equation can
be written term by term as (G\'orski 1988)
\be
\Bigl < v_{\mu}(\br_i) v_{\nu}(\br_j)\Bigr > =  \Psi_{\perp}(r_{ij})
\delta_{\mu\nu} +
[\Psi_{\Vert}(r_{ij}) - \Psi_{\perp}(r_{ij})]\hat r_{\mu} \hat r_{\nu},
\label{eq:covar}
\ee
where $\Psi_{\Vert}(r)$ and $\Psi_{\perp}(r)$ are the radial and transverse
velocity correlation functions respectively, $\br_{ij}\equiv\br_j - \br_i$, and
$\bv_{\mu}(\br_i)$ is the $\mu$ component of the peculiar velocity at $\br_i$.
These correlation functions are related to the density power spectrum via
\be
\Psi_{\perp,\Vert}(r)= {H_0^2 f(\Omega)^2\over 2 \pi^2}
\int_0^\infty P(k) K_{\perp,\Vert}(kr) \rd k ,
\label{eq:psi}
\ee
where $K_{\perp}(x) \equiv j_1(x)/ x$ and
$K_{\Vert}(x) \equiv j_0 - 2{j_1(x)/ x}$,
in which $j_l(x)$ is the spherical Bessel function of order $l$.

The cross-correlation matrix of the data and the underlying field
enters the above equations as, \eg,
\be
\Bigl < \delta(\br) \uo_j \Bigr >
=\Bigl < \delta(\br) \bv(\br_j) \Bigr > \cdot \hat \br_j .
\ee
The
two-point cross-correlation vector between the density and velocity fields
is related to the power spectrum via
\be
\Bigl < \delta(\bx)\, \bv(\bx + \br) \Bigr > =
- {H_0 f(\Omeg)\over 2  \pi^2} \hat\br
\int_0^\infty k P(k) j_1(kr) \rd k.
\label{eq:cross}
\ee

The assumption that linear theory is valid on all scales enables us to
choose the resolution as will, and in particular to use different
smoothings for the data and for the recovered fields.
In our case no smoothing were applied to the radial velocity data
while we choose to reconstruct the density field with a finite
Gaussian smoothing of radius $R$. This alters the density-velocity
correlation function by inserting the multiplicative term
$\exp[-k^2 R^2/2] $
into the integrand of \eq{cross}.

In order to estimate the {\it quality} of the reconstruction
we compute at every point a theoretical signal-to-noise ratio ($S/N$).
The minimum-variance property of the WF enables one to calculate
the variance of the residual of the actual field from the WF solution,
which provides an estimate of how much the actual field differs from
its estimator.  
The covariance of the residual of the reconstructed density field is:
\be
\Bigl< (\delta(\br) - \delta\WF(\br))\
(\delta(\br') - \delta\WF(\br'))  \Bigr> =
 \Bigl<  \delta(\br)  \delta(\br')    \Bigr> -
\Bigl < \delta(\br) u_i \Bigr >
\Bigl<   \uo_i \uo_j \Bigr >^{-1} \Bigl < \delta(\br') u_j \Bigr > .
\label{eq:cov-res}
\ee
The variance of the residual corresponds to the diagonal
terms of the covariance matrix, namely,
\be
[\sigma\WF(\br)]^2= \Bigl< [\delta(\br) - \delta\WF(\br)]^2 \Bigr>.
\ee
The $S/N$ in any point $\vr$ is then given by
\be
S/N = \left\vert {\delta\WF(\br) \over \sigma\WF(\br)} \right\vert.
\ee
In the case of random Gaussian fields, the
ensemble of CRs defined in \eq{CRv} and \eq{CRd} samples the distribution
determined by the covariance of \eq{cov-res}.
Expressions for the residual of the velocity
field can be evaluated in a similar way.

A word of caution:
at the present epoch the dynamics on small scales is definitely non-linear and
therefore the application of the linear WF/CR algorithm should be examined
with care. The two questions that arise are whether the
data and the reconstructed field both lie in the linear regime.
The maximum-likelihood analysis that was applied to the Mark III
catalog (Zaroubi \etal 1997) has
shown that for a CDM-like power spectra in the acceptable range of parameters
the linear velocity auto-correlation function is insensitive to the actual
length of smoothing on small scales. This allows us to leave the data
basically unsmoothed except of local grouping that deals with
rich clusters and helps reducing Malmquist bias.
The density field, on the other hand, has to be smoothed in order to
obey the linear approximation.

It is worth noting that the WF represents a general minimum-variance
solution under the sole assumption that the field is a random field
with a known power spectrum.
No assumption has to be made here regarding the correlations of higher
order (or the full joint probability distribution functions)
of the underlying field.
On the other hand, the CRs are derived under the explicit assumption
of a full Gaussian random field.

\section{DATA AND PRIOR POWER SPECTRUM}
\label{sec:data}

The \m3\ catalog (Willick \etal\ 1995, 1996, 1997a), which consists of more
than 3400 galaxies, has been compiled from several data sets of spirals and
elliptical/S0 galaxies with distances inferred by the forward Tully-Fisher and
$D_n-\sigma$ distance indicators.
These data were re-calibrated and self-consistently put
together as a homogeneous catalog for velocity analysis.  The catalog provides
radial velocities and inferred distances with errors on the order of
$17-21\%$ of the distance per galaxy. The sampling covers the whole sky outside
the ZoA, but with an anisotropic and non-uniform density that is a strong
function of distance. The good sampling typically ranges out to $6000\kms$ but
it may be limited to only $4000\kms$ in some directions or extend beyond
$8000\kms$ in other directions.
The ZoA is of $\pm 20-30^\circ$.

The data is carefully corrected for Malmquist biases.
A grouping procedure has been applied to the data in order
to lower the inhomogeneous Malmquist bias before correction and to
avoid strong non-linear effects (in particular large velocities of galaxies in
clusters).  This procedure yields a dataset of distances, radial peculiar
velocities and errors for $\approx 1200$ objects, ranging from individual
field galaxies to rich clusters.
In the current study we assign the errors to the
radial peculiar velocities and ignore the associated errors in the distances.
The bias that may be introduced by this procedure is small compared to the
other errors involved (\eg, Dekel, Bertschinger \& Faber 1990).
These data are used as the input for the MaxLike analysis to determine
the power spectrum and the WF reconstruction of the fields as well
as the constraints imposed on the realizations of the local structure.

The {\it prior} power spectrum has been determined in a preliminary step
by a maximum-likelihood analysis of exactly the same data (Zaroubi \etal 1997).
A parametric functional form was assumed as a model for the power spectrum
and the free parameters were determined by maximizing the probability of the
model given the data.

Under the assumption that both the underlying velocities and the
observational errors are independent Gaussian random fields, the
likelihood function that is being maximized can be written in the
following form:
\be
\label{eq:like}
{\cal L} = [ (2\pi)^N \det(R)]^{-1/2}
  \exp\left( -{1\over 2}\sum_{i,j}^N {\uo_i R_{ij}^{-1} \uo_j}\right)\,.
\ee
This is simply the corresponding multivariate Gaussian distribution.
The correlation matrix of the velocities, $R_{ij}$, can be computed
as a function of the free parameters of the power spectrum
following the computation described in \S~\ref{sec:method}, \eq{Rij}
to \eq{psi}.

This likelihood method was carefully tested using detailed mock catalogs
and found to properly reproduce unbiased power spectra.
The parametric functional forms covered the range
of CDM models with and without a cosmological constant, a tilt, and COBE
normalization. The free parameters were subsets of the following:
the density parameter $\Omega$, the cosmological parameter
$\Lambda$, the Hubble constant $H_0$ ($\equiv h\, 100\kms {\rm Mpc}^{-1}$),
the power index $n$, and the ratio of tensor to scalar fluctuations.
The results are robust (see also below in \S~\ref{sec:results}).

\section{TESTING WITH A MOCK CATALOG}
\label{sec:test}

In order to study the validity of our method and its limitations,
and to estimate the errors involved in the reconstructed fields,
we apply it first to realistic artificial catalogs that mimic
the Mark III data including all the associated uncertainties.
These mock data were ``observed" from the ``galaxy"
distribution in an N-body simulation that was pre-designed to mimic
the real structure in the local universe (as traced by the IRAS 1.2Jy redshift
survey) and thus to honestly reproduce any signal-dependent error.
These mock catalogs are described in detail in Kolatt \etal (1996)
and some features were added in Sigad \etal (1998).

We wish to investigate in particular the following three
ingredients of the method:
(1) the ability of the MaxLike analysis to estimate correctly the power
spectrum of the underlying field;
(2) the adequacy of the linear tools developed here when applied to
`observations' sampled out off an evolved non-linear field, and
(3) the variance of the residual from the WF mean field.

The true power spectrum of the mass density fluctuation field in the
simulation is well approximated by a three-parameter model of the form
\be
P(k)={A_0 k \over 1 + (B k)^\alpha }.
\label{eq:Pk}
\ee
The MaxLike analysis of Zaroubi \etal (1997), when applied to the mock data
with this functional form as a prior model, yielded values for the three
free parameters that are very close to the actual values.
%
This analysis also provides partial support to our working hypothesis that
the linear tools can be used to reconstruct the (linear) fluctuation field
from the velocities that are associated with a non-linear density field.

Given the true power spectrum as a prior, the WF/CR algorithm was
then applied to the mock \m3\ catalog with three different
smoothings. The success of the reconstructions is demonstrated via maps in
Figures 1, 2 and 3. They show maps of density and velocity fields
in the Supergalactic plane, with Gaussian smoothing of radii $R=1200$, $900$
and $500\kms$ (hereafter G12, G9 and G5 respectively).
The map at the top-left corner is the target --- the true density field
in the mock simulation (inspired by IRAS 1.2Jy). The top-right corner
shows the WF mean field.
Three different random CRs are shown, sampling the variance around the mean.
A map of the theoretical estimate of the $S/N$ is also shown.

It is seen that for the G12 smoothing the WF recovers the main features
of our local cosmography quite well, including the Great Attractor (GA)
on the left, The Perseus-Pisces supercluster (PP) on the right,
and the Local Void in between.
The theoretical $S/N$ of these is high and they are indeed
reproduced by the three CRs shown.
The WF map of smoothing G9 reproduces the
actual structures at a higher spatial resolution, but at the expense of a lower
$S/N$. The GA reconstruction is still robust but the PP region, which is
sampled more poorly, is becoming more dominated by the random component of
the CRs. This trend is naturally more pronounced
in the G5 maps. The single-peak structure of the GA as detected with
the coarse resolution now resolves into a multi-peak structure.
The WF recovers this
fine structure but with a high theoretical uncertainty. Indeed, some of
the CRs do not reproduce the same fine structure.
Compared to the GA, the reconstruction in the PP region is of lower quality,
with a lower $S/N$. The larger noise in PP manifests itself as a lower
peak in the WF density field despite the fact that the ``true" PP
in the mock simulation is actually a higher peak than the GA.
The CRs do not suffer from this problem of attenuation --- their
random components compensate for the loss of power by the WF in noisy regions.

\placefigure{fig:fig1_ZHD}

\placefigure{fig:fig2_ZHD}

\placefigure{fig:fig3_ZHD}

In summary: the WF reconstruction recovers all the main structures with
a degrading quality as a function of the attempted resolution.
As expected, attenuation affects the WF maps in the noisy regions.
The CRs recover the power and produce statistically uniform maps.
The resemblance of each CR to the ``true" field is a measure of success
of the reconstruction, compatible with the $S/N$ maps.

\section{RECONSTRUCTION FROM MARK III DATA}
\label{sec:results}

\subsection{Maps of Density and Velocity Fields}

The \prior\ model to be used by the WF reconstruction has been determined
by the MaxLike analysis of the \m3\ peculiar-velocity data
(Zaroubi \etal 1997).
The chosen mass power spectrum is of an Einstein-deSitter CDM model
($\Omega=1$ with no cosmological constant), with a large-scale tilt of $n=0.8$
and with Hubble constant $h=0.75$, normalized to fit the 4-year COBE data
on large scales. However, this power spectrum does not represent a very
specific choice of cosmological parameters, because it is almost identical
to the
power spectrum obtained by any choice of parameters that obey the empirical
relation $\Omega h^{1.2} n^2 \approx 0.45$. Thus, it is similar, for examples,
to the power-spectra of CDM models with lower $\Omega$ and less tilt,
with or without a cosmological constant.
A confirmation for the robustness of this power spectrum comes from
a recent likelihood analysis by Freudling \etal (1998)
of a complementary catalog of peculiar velocities --- the SFI catalog
by Haynes \etal (1998) --- which yields a very similar power spectra
and also confirms the error estimates.

Maps in the Supergalactic plane of the WF mean field, the theoretical $S/N$
and four different CRs are shown in Figures 4, 5 and 6, corresponding to
smoothings of G12, G9 and G5 respectively.
In the G12 maps, as in the mock data,
the structure is naturally of lower resolution
but it is characterized by a higher $S/N$ compared to the maps of higher
resolution. For example, the GA region at the left of the G12 map
is made of a single broad peak centered roughly at $(X,Y)\approx(-4000,0\kms)$,
branching out with a moderate slope towards Virgo $(-300,1300)$,
and the Local Group $(0,0)$.
The central peak of the GA shows high values of $S/N$, which are
also reflected in the relatively little scatter seen between the different
CRs near the peak region.
The density peak of the PP supercluster, on the opposite side of the sky,
shows somewhat lower values of $S/N$, reflecting the differences in sampling
in those two general regions.
Note that the $S/N$ map of Fig. 4 differs slightly from that of Fig. 1.
This is because the power spectra used in the reconstruction of the mock and
the real \m3\ catalogs were slightly different.

\placefigure{fig:fig4_ZHD}

\placefigure{fig:fig5_ZHD}

\placefigure{fig:fig6_ZHD}

In the high-resolution maps, the GA breaks into a significant multi-peak
structure, separating the central region behind Hydra and Centaurus
from the nearby structures Virgo and possibly Fornax
($Y\approx \pm 1300$ respectively).
The PP supercluster also starts showing fine structure, but of relatively
low $S/N$.

The cosmography of the G10
density field is also presented in Figure 7 by Aitoff
projections in galactic ($l,b$) coordinates at distance cuts of $R=1500$,
$3000$ and $4000\kms$.  At a distance of $1500\kms$, the two peaks that
are apparent in the map are at $(l,b)\approx(330^\circ,70^\circ)^\circ$
and $(295,-30)$, corresponding roughly to the locations of the Virgo
and Fornax clusters.  The $3000\kms$ projection
shows the bridge extending from the GA ($l\approx 300$)
to Virgo and the Local Group including the Centaurus
($l\approx 325$) and Hydra ($l\approx 280$) clusters.
The cut through the ridge that goes towards the
PP peak at $l\approx 140$ is clearly seen.
This filament starts at a
distance of about $2500\kms$ and it extends to the edge of the reconstruction
region at about $6000\kms$. The highest peak of the reconstructed
density field is seen in the $4000\kms$ projection at
$(l,b)\approx(325,-5)$, where the 'observed' redshift is $z\approx 4400\kms$.
This peak closely matches the very rich cluster A3627
that was discovered at $(l,b,z)=(325^\circ, -7^\circ, 4882\kms )$
by Kraan-Korteweg \etal (1996) in a deep ZoA search near the location
predicted for the ``center" of the GA by the POTENT
analysis of Mark III (Kolatt, Dekel \& Lahav 199x).
The redshift difference can be due to smoothing and possibly nonlinear
effects.

\placefigure{fig:fig7_ZHD}

The low resolution WF maps are in general very similar to the maps obtained by
POTENT reconstruction from the same data (Dekel \etal 1998).
The same big structures of GA, PP and the void between them
are recovered at similar positions, with similar structure and extent
and with comparable density contrasts. The interpolation
into the ZoA and into large radii is safer with the WF method;
noisy features that show up in the POTENT maps in these regions
are not present in the WF maps.
The $S/N$ map, and the CRs, then provide important information about the
quality of the reconstruction, which complements the error
maps as derived from POTENT reconstructions of mock catalogs.
The high-resolution WF maps recover new substructures that do not show
up in the POTENT G12 maps. This is a result of the noise-determined
variable smoothing that is intrinsic to the WF reconstruction.
The POTENT reconstruction, although also capable of variable smoothing,
has so far been deliberately applied with fixed smoothing to ensure
spatial uniformity of the reconstruction.  When POTENT is applied with
fixed, high resolution, it amplifies specific noisy features in the
poorly sampled regions.  With the CRs, we get instead random variations in
these noisy regions and can easilly identify them.

\subsection{Comparison with IRAS and Velocity Decomposition}

A comparison of the peculiar-velocity data with redshift-survey data
is very interesting for many reasons, ranging from the need to investigate
systematic errors in the two kinds of data to the desire to measure
the cosmological parameters to understand galaxy formation.
This comparison has been tried already in several ways,
always yielding gross agreement about the main structures but sometimes
being un-decisive about the fine details of the comparison and the quality of
the agreement (\eg, recently, Davis, Nusser, \& Willick 1996;
Willick \etal 199x; Sigad \etal 1998).

We carry this comparison here via WF mean fields.
Kolatt \etal (1996) provide a nonlinear reconstruction of the IRAS 1.2Jy
redshift survey using a multi-stage procedure involving dynamical evolution
backwards and forward in time, Gaussianization and constrained realizations.
They then generate from the reconstructed density field
mock catalogs of the Mark III catalog. We apply the WF procedure to
these mock catalogs and obtain a WF mean field of the IRAS data
to be directly compared with the WF mean field as obtained in this paper
from the real Mark III peculiar velocities.
The compared fields suffer from the same statistical errors and therefore
from the same attenuation that is imposed by the WF given these errors.
The comparison carried out here is not yet fully quantitative;
it is only meant to identify gross systematic discrepancies in the
cosmography, and to present a new way of comparison to be implemented
in detail in a subsequent paper.

The recovered WF fields of Mark III and IRAS 1.2Jy are shown in Figure 8.
The velocity field is represented by projected velocity vectors, as usual,
but also by corresponding flow lines, in order to emphasize the
large-scale flows.
The comparison shows a general agreement between the two density fields.
The main difference in the density field lies in the structure of the
GA complex, in which the IRAS reconstruction puts the GA density peak
at roughly $(X,Y)\approx (-3000,1000)$ compared with the
\m3\ GA peak at $(X,Y)\approx (-4000,0)$. The GA peak has a high
$S/N$ which makes this difference significant.
We note that a similar discrepancy in the peak of the GA is apparent in the
density-density comparison of Sigad \etal (1998), leaving this region out
of the quantitative comparison done there.

The two velocity fields seem at first to be quite different.
However, one should recall that the Mark III data measure the full velocity
field while the IRAS prediction for the velocity field is derived from
the particular solution of the Poisson equation within a given
volume and 
boundary conditions, and is missing the homogenous solution,
namely the tidal field due to the mass distribution outside this volume.
To improve the comparison we therefore {\it decompose\,} the \m3\ WF
velocity field into its divergent and tidal components 
relative to a cube of $\pm 8000 \kms$ centered on the Local Group.
The divergent component is obtained by the particular solution of the
Poisson equation given the WF-reconstructed density field within this
volume.
In practice, this is done by zero-padding the WF density field
in a box twice as large on the side and then solving the Poisson equation
by FFT.
The tidal component is the residual left by subtructing the divergent
component from the full velocity.

\placefigure{fig:fig8_ZHD}

The resulting divergent and tidal fields are also shown in Fig. 8.
Indeed, the divergent component matches the IRAS velocity field much better
than the total field. The tidal field is dominated by a
bulk flow of $194\pm 32\kms$
that runs across the local neighborhood roughly in the direction
of the Shapely concentration (Bardelli \etal\ 1996 and references therein).
There is also an indication for a quadrupole component in the tidal field,
whose major axis lies in a similar direction.
A more quantitative analysis of this velocity-field decomposition
is deferred to a subsequent paper.

\section{CONCLUSION AND DISCUSSION}
\label{sec:conclusion}

The main purpose of this paper was to present an alternative method
of reconstruction from peculiar velocity data.
The WF/CR method stresses a more rigorous treatment of the random errors
than the POTENT method,
and it allows extrapolation into poorly sampled regions.
In this first paper we have mainly set up the basic formalism
and demonstrated how the method works.
We then carried out basic tests using mock catalogs,
and recovered the underlying density and velocity fields from the
Mark III data.
Future papers will describe more quantitative analysis based on this method.

The results reported here are limited to maps of different resolution
in the Supergalactic plane, and corresponding error maps.
The WF mean field is our best bet for the underlying fields given the
data and the errors. It represents a compromise between the constraints
imposed by the data and the \prior\ model. As a result, the statistical
properties of the WF field are not spatially uniform --- the density
field suffers from an artificial attenuation towards the mean of the
prior, $\delta=0$, where the noise is dominant.
The CRs compensate for this effect, and produce random realizations which
are statistically uniform in space. Each of these realizations could be
the best approximation to the real cosmological neighborhood with equal
probability.  In the well-sampled regions and wavelengths the CRs are
strongly constrained by the data, and where the noise dominates they
approach the limit of unconstrained random realizations of the prior model.
The variations among the CRs thus provide an interesting
representation of the uncertainty in the reconstruction, and the CRs provide
clear indications for the robustness of the various structures.

The low resolution field is dominated by the data in much of the volume
shown, and is hardly affected by the \prior\ model. The resulted density
field is very similar to the one produced by the direct POTENT algorithm.
The WF map in the inner regions is characterized by a high $S/N$,
and there are relatively little variations between the different CRs.
In the WF maps of higher resolution, small-scale structures are resolved
which are not clearly identified in the raw data.  The properties of this fine
structure depend on the nature of the \prior\ model, and so is the theoretical
$S/N$. The CRs exhibit high variations, consistent with the low $S/N$.
It is clear that the WF extrapolation in k-space is valid only over a limited
dynamical range; in the limit of very high resolution one expects
a null WF density map and unconstrained realizations.

A useful feature of the WF/CR reconstruction method is that it can
relate two datasets that differ in many ways.
For example, the fact that it can translate velocities into densities
allows a comparison of velocity and density data, even if sampled with
different resolutions and not at exactly the same positions.
This is the mode of operation used here to qualitatively
compare the \m3\ data with the IRAS 1.2Jy data. The method can alternatively
be used to create realistic mock catalogs of actual datasets, such as the mock
\m3\ catalog made of the non-linear constrained realization of the IRAS
density field (Kolatt \etal\ 1996).

An apparent difficulty arises from the fact that
the current WF/CR reconstruction assumes linear gravitational
instability, yet it is applied to a universe that is
{\it non-linear} on scales smaller than a few megaparsecs.
Based on our tests with Mock catalogs (\eg, Zaroubi \etal 1997), we
can assume that the grouped galaxy velocities indeed serve as reasonable
tracers of the linear velocity field.
But our current procedure carries the application of linear theory to
an extreme by assuming that the present-day density field obeys linear theory
even on small scales.
To obtain the non-linear structure on these small scales, the WF/CR field
can be taken back in time and used as initial conditions for an $N$-body
simulation that then evolves the gravitating system forward in time in a
fully nonlinear way (see Kolatt \etal 1996; Bistolas and Hoffman 1998).
A quantitative comparison with observed nonlinear structure would
require such a procedure.  For example, the identification of the
highest density peak in the linear reconstruction with
the cluster A3627 is valid only in the context of identifying linear
peaks as the progenitors of final clusters.

The WF field is very suitable for the decomposition of the recovered
velocity field into its divergent and tidal components relative to a
certain, relatively large volume. This is because the WF allows a
reliable recovery of the density field even in regions that are sampled
quite sparsely.
In the language of density fields, the tidal field is
the solution of the source-free Laplace equation which depends on the
boundary conditions.  Therefore, the analysis of the tidal field,
as determined by the boundary conditions of a given peculiar velocities 
data set, can shed
light on the structure beyond the actual surveyed region.
The decomposition provides the means for a self-consistent comparison
of the velocity field reconstructed from radial
velocities with the one predicted from the density of galaxies
in a redshift survey, and for determining the cosmological parameter
$\beta=\Omega^{0.6}/b$ from such a velocity-velocity comparison.
The velocity fields from Mark III and from IRAS 1.2Jy,
which seem to be different when compared as are, become very similar
when their divergent components are compared.
A detailed analysis of the decomposition of the velocity field is
deferred to a subsequent paper.

\acknowledgments{
We gratefully acknowledge the
hospitality of the Center for Particle Astrophysics and the Astronomy
Department of UC Berkeley and Kapteyn astronomical Institute.
This work is supported in part by the US-Israel
Binational Science Foundation grants 95-00330 (AD) and 94-00185 (YH), the
Israel Science Foundation grants 950/95 (AD) and 590/94 (YH), and the
S.A. Schonbrunn Research Endowment Fund (YH).}

{}



\clearpage

\bigskip
\cl{\bf Figure Captions}
\bigskip

\figcaption[fig1_ZHD.ps]
{
Testing the method with mock Mark III data.
Shown are maps of density and velocity fields in the Supergalactic plane.
The smoothing window is a Gaussian of radius $1200\kms$, G12.
Density contour spacing is $0.1$, the mean $\delta=0$ contour is heavy,
positive contours are solid and negative contours are dashed.
The velocity field is presented by arrows with arbitrary scaling.
Top-left: The target ``true" density field of the simulation.
Top-right: The reconstructed Wiener Filter mean field.
Middle-left: The theoretical signal--to--noise ratio with
contour spacing 1.  The shading indicates
regions where the error is less then 0.36.  
The three panels at the bottom-right show three different constrained
realizations. The velocity field is added only at the bottom-right panel.
\label{fig:fig1_ZHD}
}

\figcaption[fig2_ZHD.ps]
{
Same as Fig. 1, but with G9 smoothing. The density contour spacing is
now $0.15$, except in the bottom-right panel where it is $0.3$.
The $S/N$ spacing is 1, and the shading refers to error smaller than
0.45. 
\label{fig:fig2_ZHD}
}

\figcaption[fig3_ZHD.ps]
{
Same as Fig. 1, but with G5 smoothing. The density contour spacing is
now $0.2$, except in the bottom-right panel where it is $0.5$.
The $S/N$ spacing is 1, and the shading refers to error smaller than
0.63. 
\label{fig:fig3_ZHD}
}

\figcaption[fig4_ZHD.ps]
{
Reconstruction from the Mark III data.
Shown are maps of density and velocity fields in the Supergalactic plane.
The smoothing window is G12.  Contours are as in Fig. 1.
Top-left: The reconstructed Wiener Filter mean field.
Middle-left: The theoretical signal--to--noise ratio with
contour spacing 1.  The shading indicates
regions where the error is less then 0.41.  
The four other panels show four different constrained
realizations. The velocity field is added only at the bottom panels.
\label{fig:fig4_ZHD}
}

\figcaption[fig5_ZHD.ps]
{
Same as Fig. 4, but with a G9 smoothing. The density contour spacing is
now $0.15$, except in the bottom panels where it is $0.3$.
The $S/N$ spacing is 1 and the shading refers to error smaller than 0.49.
\label{fig:fig5_ZHD}
}

\figcaption[fig6_ZHD.ps]
{
Same as Fig. 4, but with a G9 smoothing. The density contour spacing is
now $0.2$, except in the bottom panels where it is $0.5$.
The $S/N$ spacing is 1 and the shading refers to error smaller than 0.71.
\label{fig:fig6_ZHD}
}

\figcaption[fig7_ZHD.ps]
{
The reconstructed WF density field, smoothed G10,
shown on spherical shells at distances 1500, 3000 and $4000\kms$.
The Aitoff projection is shown in Galactic coordinates ($l,b$).
The Supergalactic plane is marked by filled circles.
The density contour spacing is 0.2, with contours as in the
previous figures.
\label{fig:fig7_ZHD}
}

\figcaption[fig8_ZHD.ps]
{
A comparison of the WF mean fields in the Supergalactic plane, smoothed
G5, as reconstructed from the Mark III velocities and from the IRAS 1.2Jy
redshift survey.
Top: density maps, IRAS on the left and Mark III on the right.
Contours are as in the previous figures, with spacing 0.1.
The velocity fields are displayed as flow lines that start at random
points, continue tangent to the local velocity field, and are of
length proportional to the magnitude of the velocity at the starting
point.
Middle: velocity fields, IRAS on the left and Mark III on the right.
Bottom: The WF from Mark III is decomposed into its
divergent (left) and tidal (right) components, with respect to a cube
of $\pm 8000\kms$ about the Local Group.
\label{fig:fig8_ZHD}
}

\end{document}